\documentclass[aps,pra,reprint,superscriptaddress,floatfix]{revtex4-2}
\usepackage{amsmath,amssymb}
\usepackage{graphicx}
\graphicspath{{./}}
\usepackage{booktabs}
\usepackage[hidelinks]{hyperref}
\begin{document}

\title{Phase-conjugate Gaussian encoding for continuous-variable quantum key distribution with a single phase-sensitive amplifier}
\author{Zhenlin Zhao}
\affiliation{School of Electronics and Information Technology, Sun Yat-sen University, Guangzhou, China}
\author{Dawei Wang}
\email{wangdw9@mail.sysu.edu.cn}
\affiliation{School of Electronics and Information Technology, Sun Yat-sen University, Guangzhou, China}

\begin{abstract}
Phase-conjugate encoding is naturally compatible with shared phase-sensitive amplification and joint detection, offering a compact optical receiver architecture for continuous-variable quantum key distribution (CVQKD).
Here, we develop a unified encoding and security framework for two-mode Gaussian CVQKD with tunable phase-conjugate intermode correlations matched to a joint receiver comprising a single phase-sensitive amplifier (PSA) and one balanced photodetector.
We present both coherent-state and squeezed-state prepare-and-measure protocols that use the same transmitted ensemble.
The two preparations can be implemented through classically precoded Gaussian displacements of coherent sidebands or a directly generated two-mode squeezed vacuum, respectively, without an actual optical mode converter.
The encoding and detection scheme separates low-variance joint quadratures for key generation from high-variance joint quadratures carrying private random modulation.
This complementary modulation can reduce Eve's Holevo information without changing the Alice--Bob mutual information.
Numerical results show that optimized correlated coherent-state encoding achieves higher asymptotic key rates at long distances and extends the positive-key range compared with optimized single-mode coherent-state modulation.
The optimal intermode correlation increases with transmission distance, and the range advantage persists under the symmetric correlated two-mode Gaussian attacks considered here.
These results provide a theoretical basis for the design and implementation of PSA-enhanced two-mode CVQKD with phase-conjugate encoding.
\end{abstract}

\maketitle

\section{Introduction}
\label{sec:introduction}

Quantum key distribution (QKD) enables two distant parties, Alice and Bob, to establish secret keys using quantum states and authenticated classical communication \cite{scarani2009security,pirandola2020advances}.
Continuous-variable (CV) QKD encodes information in the quadratures of optical fields and can be implemented with Gaussian-modulated coherent or squeezed states and coherent receivers \cite{grosshans2002continuous,weedbrook2012gaussian}.
The intrinsic quantum state and information encoding used to prepare Alice's ensemble are defining features of CVQKD protocols and strongly influence their achievable performance \cite{cerf2001quantum,grosshans2002continuous,leverrier2009unconditional,weedbrook2010quantum,usenko2015unidimensional,jacobsen2018complete,winnel2021minimization,nguyen2023imbalanced,erkilic2025enhanced}.
Although most CVQKD protocols use single-mode encoding, multimode signals with engineered intermode correlations allow information to be encoded in collective quadratures and accessed through joint measurements \cite{usenko2014multimode}, which can be realized with a compact optical receiver.

A concrete example is phase-conjugate two-mode encoding, in which two real modulation variables $X$ and $P$ are mapped onto the coherent-state pair $|\alpha\rangle_1|\alpha^*\rangle_2$, with $\alpha=(X+iP)/2$ in shot-noise units.
The two modes therefore carry displacements $(X,P)$ and $(X,-P)$, respectively.
The encoded information can be measured simultaneously through the collective quadratures $\hat{x}_+=(\hat{x}_1+\hat{x}_2)/\sqrt{2}$ and $\hat{p}_-=(\hat{p}_1-\hat{p}_2)/\sqrt{2}$, which commute and have mean values of $\sqrt{2}X$ and $\sqrt{2}P$, respectively.
Cerf and Iblisdir showed that jointly measuring this phase-conjugate pair provides more information about the encoded displacements than using two identical coherent-state copies \cite{cerf2001phase}.
In their proposed realization, a balanced beam splitter maps $\hat{x}_+$ and $\hat{p}_-$ onto complementary quadratures of two output modes, which are then measured by homodyne detection.
Niset \textit{et al.} subsequently proved and experimentally demonstrated that joint measurements on Gaussian-distributed phase-conjugate coherent-state pairs reconstruct $|\alpha\rangle$ with higher fidelity than any strategy that measures the two modes separately, even when the local operations are coordinated through classical communication \cite{niset2007superiority}.
Arnhem \textit{et al.} further established the optimality of phase-conjugate coherent-state encoding for simultaneous estimation of $X$ and $P$ under a global energy constraint.
They showed that phase-conjugate encoding with joint detection simultaneously attains the quantum Cram\'er--Rao bounds for both variables \cite{arnhem2019optimal}.
In CVQKD, Su \textit{et al.} demonstrated phase-conjugated subcarrier modulation and phase-sensitive joint detection using a single balanced photodetector, connecting correlated Gaussian encoding to compact joint optical reception \cite{su2023experimental}.

However, the practical benefit of this joint measurement is limited by finite detection efficiency and electronic noise, which reduce the useful correlations available for reconciliation.
Receiver-side optical preamplification can mitigate these detection penalties by amplifying the signal before photodetection \cite{fossier2009improvement,shaked2018lifting}.
A phase-sensitive amplifier (PSA) amplifies selected quadratures without the added-noise requirement of phase-insensitive amplification \cite{caves1982quantum}.
Phase-sensitive amplification was considered in early squeezed-state quantum cryptography \cite{hillery2000quantum} and has subsequently been investigated for Gaussian coherent-state protocols \cite{fossier2009improvement}, discrete modulation \cite{zhang2012improving,guo2018performance}, two-way communication \cite{zhang2014improvement}, unidimensional modulation \cite{huang2019unidimensional}, and self-referenced CVQKD \cite{qi2021selfreferenced}.

Phase-conjugate encoding offers a natural route to combining phase-sensitive preamplification with joint detection, since a single two-mode PSA can amplify a selected pair of commuting collective quadratures that match the observables accessed by the joint receiver \cite{tong2011ultrasensitive}.
For phase-conjugate frequency sidebands, these quadratures can be read out using a central local oscillator (LO) and one balanced photodetector (BPD), followed by electrical quadrature demodulation, as demonstrated experimentally in PSA-enhanced high-speed CVQKD \cite{liao2025experimental}.
The compatibility between paired encoding, amplification, and joint detection motivates a unified analysis of how the transmitted intermode correlations and Alice's quantum-state preparation affect the achievable security and performance.

In this work, we consider tunable phase-conjugate Gaussian encoding matched to a joint receiver with a single phase sensitive preamplifier and construct both coherent-state and squeezed-state preparations from a common entanglement-based representation.
We relate the encoding to independently prepared asymmetric ensembles and show that the two-mode encoding is practically more appealing because the mode conversion occurs at the source, avoiding physical mode conversion before the joint receiver.
Moreover, using coherent states, Alice's implementation is further simplified by directly generating the correlated sidebands through classical precoding and modulation.
Using a trusted PSA-enhanced receiver characterized by its net gain and noise figure, we compare correlated coherent-state encoding with single-mode coherent-state modulation under the same receiver-noise conditions.
Numerical results show that the single-mode benchmark achieves higher rates at short and intermediate distances, while correlated encoding gives higher rates at long distances and extends the positive-key range.
We further examine performance under correlated two-mode Gaussian attacks and find that the advantage in maximum transmission distance over the single-mode coherent-state benchmark persists.
Together with the analysis of detector imperfections and mode-conversion loss, these results provide a theoretical basis for the design and implementation of practical two-mode CVQKD with phase-conjugate encoding and PSA-enhanced joint reception.

\begin{figure}[!t]
\centering
\includegraphics[width=\columnwidth]{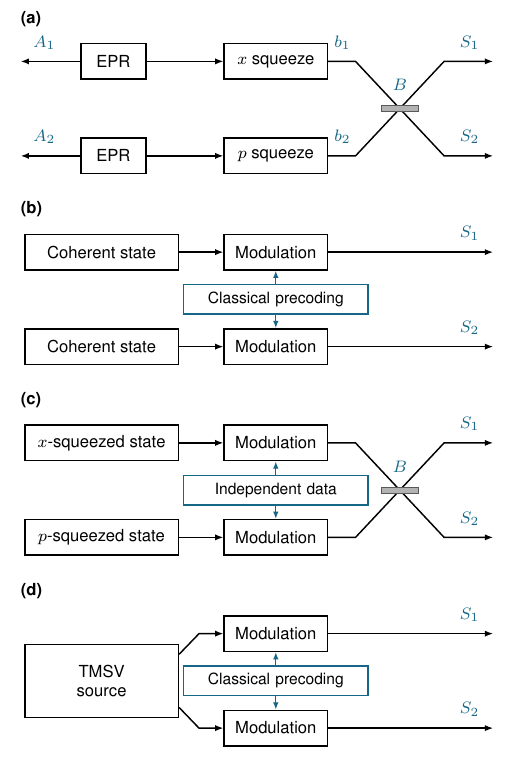}
\caption{The EB source and two equivalent preparations.
(a) The purification of $\Gamma_{S_1,S_2}$ with EPR and asymmetric squeezing.
(b) Coherent-state preparation uses classical precoding and direct modulation, no optical mixer required.
(c) Oppositely squeezed states are independently Gaussian-modulated and mixed by $B$.
(d) Equivalently, a TMSV source is followed by correlated modulation driven by classical precoding.}
\label{fig:common_eb}
\end{figure}

\section{Phase-Conjugate Gaussian Encoding}
\label{sec:source}

Consider a zero-mean two-mode Gaussian ensemble with tunable intermode correlation specified by a nonnegative parameter $c$.
Write the covariance matrix (CM) of this ensemble in physical-mode ordering $(x_1,p_1,x_2,p_2)$ as
\begin{equation}
\Gamma_{S_1,S_2}=
\begin{pmatrix}
VI_2&cZ\\
cZ&VI_2
\end{pmatrix},
\label{eq:source}
\end{equation}
where $I_2$ is the identity matrix and $Z=\operatorname{diag}(1,-1)$.
We restrict the family to $V>1$ and $0\leq c\leq V-1$ \cite{simon2000peres}; the two endpoints correspond to uncorrelated modes and maximal classical phase-conjugate correlation, respectively.

A common entanglement-based (EB) source provides a purification of ~\eqref{eq:source}, as illustrated in Fig.~\ref{fig:common_eb}(a).
Alice starts with two independent pure EPR pairs, each with variance $u = \sqrt{V^2-c^2}$.
She holds one mode of each pair, denoted by $A_1$ and $A_2$, and applies opposite squeezing $S_r=\operatorname{diag}(e^{-r},e^{r})$, and $S_{-r}=\operatorname{diag}(e^{r},e^{-r})$ to the outgoing modes, where
\begin{equation}
r=\frac14\ln\frac{V+c}{V-c}
\ .
\end{equation}
Alice then mixes these two modes $b_1$ and $b_2$ using the balanced transformation
\begin{equation}
B=\frac{1}{\sqrt2}
\begin{pmatrix}I_2&I_2\\-I_2&I_2\end{pmatrix},
\label{eq:bs}
\end{equation}
producing the signal modes $S_1, S_2$ with CM \eqref{eq:source}.
The four-mode state $\{A_1, A_2, S_1, S_2\}$ is pure and constitutes the EB representation of the source.
The full purification is detailed in Appendix~\ref{app:decompositions}.

The domain $0\leq c\leq V-1$ admits both coherent-state and squeezed-state preparations from the same EB source, as described below.

\subsection{Coherent-state preparation}
\label{sec:coherent_preparation}

When Alice performs a generalized heterodyne measurement on mode $A_1$ of the EB source, characterized by the equivalent measurement noise CM
\begin{equation}
\begin{gathered}
M=\operatorname{diag}(m,m^{-1}),
\qquad
m=\frac{u-s}{V-c-1}
\ ,
\end{gathered}
\label{eq:coherent_seed}
\end{equation}
and $s = e^{-2r}$, she remotely prepares a coherent state in $b_1$, as derived in Appendix~\ref{app:decompositions}.
The measurement on $A_2$ uses the exchanged entries $\operatorname{diag}(m^{-1},m)$ and likewise prepares a coherent state in $b_2$.
The average CMs of $b_1$ and $b_2$ are $\operatorname{diag}(V-c,V+c)$ and $\operatorname{diag}(V+c,V-c)$, respectively.
They describe two independently prepared, oppositely oriented asymmetric coherent-state ensembles.
Finally, the balanced transformation $B$ introduces the correlation in \eqref{eq:source}.

Since a passive beam splitter maps a product of coherent states to another product of coherent states, Alice can implement the physical mode transformation directly through classical precoding.
An explicit realization goes as follows: Alice uses independent zero-mean Gaussian variables $X, P$ of variance $c$, together with four mutually independent Gaussian variables $n_{1x},n_{2x},n_{1p},n_{2p}$ of variance $\varepsilon = V-c-1$, which are also independent of $X, P$.
Alice prepares the product state $|\alpha_1\rangle\otimes|\alpha_2\rangle$ with
\begin{align}
\overline x_1&=X+n_{1x},&
\overline p_1&=P+n_{1p},\nonumber\\
\overline x_2&=X+n_{2x},&
\overline p_2&=-P+n_{2p},
\label{eq:coherent_modulation}
\end{align}
and $\alpha_j=(\overline x_j+i\overline p_j)/2$.
When $\varepsilon=0$, the two amplitudes become exact phase conjugates, corresponding to the boundary discussed in Sec.~\ref{sec:encoding_boundaries}.
For $\varepsilon>0$, individual amplitudes are no longer exact conjugates, and the ensemble has the tunable phase-conjugate correlation.

In practice, Alice can synthesize the required displacements \eqref{eq:coherent_modulation} directly on two coherent-state sources, such as two sidebands of a single-mode laser, as illustrated in Fig.~\ref{fig:common_eb}(b).

\subsection{Squeezed-state preparation}
\label{sec:squeezed_preparation}

For the same EB source, Alice can instead perform ordinary heterodyne measurements on both her modes with measurement-noise CM $M=I_2$ and remotely prepare two squeezed states in $b_1$ and $b_2$ with CMs $\operatorname{diag}(s,s^{-1})$ and $\operatorname{diag}(s^{-1},s)$, respectively.
The average CMs of $b_1$ and $b_2$ are exactly the same as in the coherent preparation.

In the PM implementation, Alice prepares two oppositely squeezed vacuum states, applies required Gaussian displacements, and mixes the modes with $B$, as illustrated in Fig.~\ref{fig:common_eb}(c).
The intrinsic outgoing state is therefore a two-mode squeezed vacuum (TMSV) with CM
\begin{equation}
\Gamma_*=
\begin{pmatrix}
\cosh(2r)I_2&\sinh(2r)Z\\
\sinh(2r)Z&\cosh(2r)I_2
\end{pmatrix}.
\label{eq:intrinsic_tmsv}
\end{equation}

An equivalent implementation can directly generate the TMSV through a suitable nonlinear source and then add correlated displacement accordingly \cite{weedbrook2012gaussian}; in this case, Alice also uses only classical precoding, similar to the coherent-state protocol above, but the single laser source must be replaced by a TMSV source, as illustrated in Fig.~\ref{fig:common_eb}(d).

\begin{table}[!t]
\caption{PM decompositions of the same ensemble.
Intrinsic and modulation variances are in $(x,p)$ order for $b_1$ before mixing; $b_2$ exchanges the entries.}
\label{tab:pm_decompositions}
\centering
\small
\setlength{\tabcolsep}{3pt}
\renewcommand{\arraystretch}{1}
\begin{tabular*}{\columnwidth}{@{\extracolsep{\fill}}ccccc@{}}
\toprule
$\rho$ & PM & $M_1$ & Intrinsic & Modulation \\
\midrule
$[0,1]$ & Coh. & $\begin{bmatrix}m&0\\0&m^{-1}\end{bmatrix}$ &
$\begin{bmatrix}1\\1\end{bmatrix}$ &
$\begin{bmatrix}V-c-1\\V+c-1\end{bmatrix}$ \\
\addlinespace[1pt]
$0$ & Coh. & $I_2$ &
$\begin{bmatrix}1\\1\end{bmatrix}$ &
$\begin{bmatrix}V-1\\V-1\end{bmatrix}$ \\
\addlinespace[1pt]
$1$ & Coh. & $\begin{bmatrix}\infty&0\\0&0\end{bmatrix}$ &
$\begin{bmatrix}1\\1\end{bmatrix}$ &
$\begin{bmatrix}0\\2V-2\end{bmatrix}$ \\
\midrule
$[0,1]$ & Sq. & $I_2$ &
$\begin{bmatrix}s\\s^{-1}\end{bmatrix}$ &
$\begin{bmatrix}V-c-s\\V+c-s^{-1}\end{bmatrix}$ \\
\addlinespace[1pt]
$0$ & Sq. & $I_2$ &
$\begin{bmatrix}1\\1\end{bmatrix}$ &
$\begin{bmatrix}V-1\\V-1\end{bmatrix}$ \\
\addlinespace[1pt]
$1$ & Sq. & $I_2$ &
$\begin{bmatrix}u^{-1}\\u\end{bmatrix}$ &
$\begin{bmatrix}1-u^{-1}\\u^2-u\end{bmatrix}$ \\
\bottomrule
\end{tabular*}
\par\smallskip
\raggedright\footnotesize
Coh.: coherent; Sq.: squeezed.
$M_1$ is the measurement-noise CM on $A_1$; $A_2$ exchanges its diagonal entries.
See Eq.~\eqref{eq:coherent_seed} for $m$.
$u = \sqrt{V^2-c^2}$ and $s = e^{-2r}$.
The coherent preparation at $\rho=1$ is understood as the $m\to\infty$ homodyne limit.
At $\rho=0$, the two preparations coincide.
\par
\end{table}

\subsection{Boundary cases}
\label{sec:encoding_boundaries}

At $c=0$, one has $u=V$, and $s=1$.
The two preparations coincide and produce the physical ensemble of two independent thermal Gaussian states with CM $\Gamma_0 = VI_4$, which is the source for two parallel single-mode coherent-state GG02-like protocols \cite{grosshans2003gaussian}.

At $c=V-1$, one has $\varepsilon=0$, and the coherent-state implementation consists of exact phase-conjugate pairs
\begin{equation}
|\psi_{X,P}\rangle=|\alpha\rangle_1\otimes|\alpha^*\rangle_2,
\qquad \alpha=\frac{X+iP}{2},
\label{eq:conjugate_coherent_pair}
\end{equation}
where $X$ and $P$ are independent zero-mean Gaussian variables of variance $V_A=V-1$ known to Alice.
The two-mode ensemble has maximal classical correlation:
\begin{equation}
\Gamma_1 =
\begin{pmatrix}
(V_A+1)I_2&V_AZ\\
V_AZ&(V_A+1)I_2
\end{pmatrix}.
\label{eq:exact_conjugate_average}
\end{equation}
The average CMs of two modes before mode mixing are $\operatorname{diag}(1,2V-1)$ and $\operatorname{diag}(2V-1,1)$, which is a pair of oppositely oriented unidimensional modulations \cite{usenko2015unidimensional}.
In this case, coherent-state protocols must deal with the estimation problem of unmodulated quadratures.

For squeezed preparation at the same boundary, the source becomes the one discussed in the PCTB minimum-leakage protocol \cite{zhao2026deterministic}, where
\begin{equation}
\begin{gathered}
u=\sqrt{2V-1},\qquad
s=\frac{1}{\sqrt{2V-1}}.
\end{gathered}
\label{eq:encoding_minimum_leakage}
\end{equation}
The squeezed decomposition modulates all quadratures and therefore has no unmodulated directions of the kind encountered in the coherent decomposition above.

Define a normalized correlation strength
\begin{equation}
\rho = \frac{c}{V-1},
\qquad
0\le\rho\le 1
\ .
\label{eq:rho}
\end{equation}
Table~\ref{tab:pm_decompositions} summarizes the two PM decompositions throughout the allowed correlation range.

\begin{figure*}[!t]
\centering
\includegraphics[width=\textwidth]{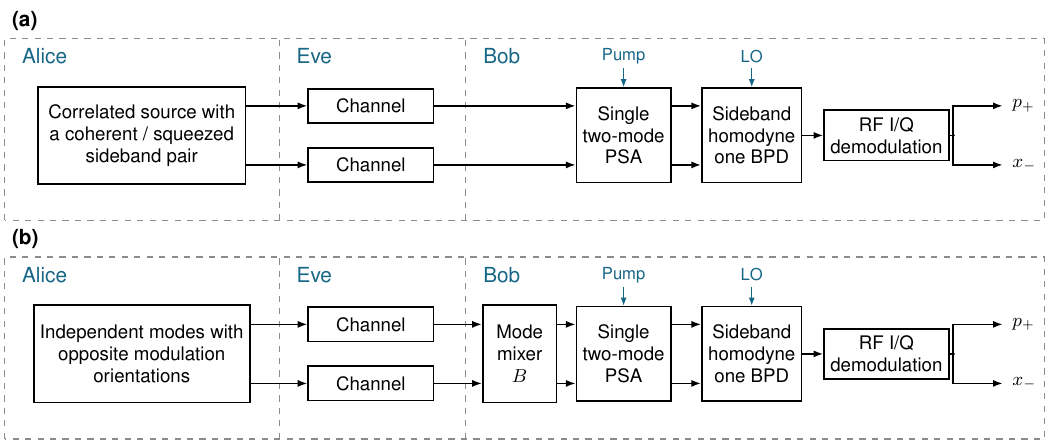}
\caption{Phase-conjugate encoding and an equivalent implementation.
(a) Alice directly prepares the correlated sideband pair and Bob uses one two-mode PSA, a central LO, and one BPD with RF I/Q demodulation.
(b) Independently prepared, oppositely oriented asymmetric modes undergo the balanced transformation $B$ before the same joint receiver.
For identical independent phase-insensitive channels and ideal mixing, the two routes have the same per-mode information quantities.}
\label{fig:protocol}
\end{figure*}

\section{Joint Reception and Security Model}
\label{sec:joint_reception}
\label{sec:security}

To connect the source encoding to Bob's receiver, we introduce the joint quadrature operators
\begin{equation}
\hat{x}_\pm=\frac{\hat{x}_1\pm\hat{x}_2}{\sqrt2},
\qquad
\hat{p}_\pm=\frac{\hat{p}_1\pm\hat{p}_2}{\sqrt2}.
\label{eq:joint_quadratures}
\end{equation}
With $[\hat{x}_j,\hat{p}_k]=2i\delta_{jk}$, we have two pairs of commuting observables $[\hat{x}_-,\hat{p}_+]=[\hat{x}_+,\hat{p}_-]=0$, and each pair can be measured jointly.
The transformation in \eqref{eq:joint_quadratures} is simply $B^T$ with $B$ given in \eqref{eq:bs}.
Namely,
\begin{equation}
B^T
\begin{pmatrix}x_1\\p_1\\x_2\\p_2\end{pmatrix}
=
\begin{pmatrix}x_-\\p_-\\x_+\\p_+\end{pmatrix}.
\end{equation}

For the source described by \eqref{eq:source}, the joint quadrature variances are
\begin{equation}
\begin{aligned}
\operatorname{Var}(\hat{x}_-)&=\operatorname{Var}(\hat{p}_+)=V-c,\\
\operatorname{Var}(\hat{x}_+)&=\operatorname{Var}(\hat{p}_-)=V+c.
\end{aligned}
\label{eq:joint_variances}
\end{equation}
The mode correlation $c$ thus allocates different variances to two commuting pairs.
Unless otherwise stated, the low-variance pair $(x_-,p_+)$ supplies the key data, while the complementary pair $(x_+,p_-)$ is used for parameter estimation.
The quadrature selection rule is motivated by the numerical results on system performance detailed in Section \ref{sec:results}.

For the coherent preparation in \eqref{eq:coherent_modulation}, Alice's corresponding records are the displacements of the joint quadratures $(x_-,p_+)$, namely,
\begin{equation}
\begin{aligned}
a_x&=\frac{\overline{x}_1-\overline{x}_2}{\sqrt2}
=\frac{n_{1x}-n_{2x}}{\sqrt2},\\
a_p&=\frac{\overline{p}_1+\overline{p}_2}{\sqrt2}
=\frac{n_{1p}+n_{2p}}{\sqrt2}.
\end{aligned}
\label{eq:coherent_key_records}
\end{equation}
Each has variance $\varepsilon = V-c-1$.
The common variables $X$ and $P$ cancel from these records and instead contribute to the high-variance pair.
For the squeezed preparation, the displacement variance is $\varepsilon = V-c-s$.

At the receiver, Bob sets the pump phase of an ideal two-mode PSA to amplify $(x_-,p_+)$.
Its symplectic transformation in the physical-mode ordering is \cite{weedbrook2012gaussian}
\begin{equation}
\begin{gathered}
S_G=\begin{pmatrix}\mu I_2&-\nu Z\\-\nu Z&\mu I_2\end{pmatrix},\\
\mu=\frac{\sqrt G+ \sqrt G^{-1}}2,\qquad
\nu=\frac{\sqrt G- \sqrt G^{-1}}2,
\end{gathered}
\label{eq:ideal_psa}
\end{equation}
where $G\geq1$ is the power gain.
In the joint-quadrature basis $(x_-,p_-,x_+,p_+)$, the transformation is
\begin{equation}
B^TS_GB=\operatorname{diag}(\sqrt G,1/\sqrt G,1/\sqrt G,\sqrt G).
\label{eq:normal_psa}
\end{equation}
The PSA thus amplifies both commuting key quadratures $(x_-,p_+)$ by $\sqrt G$ and deamplifies the complementary pair by $1/\sqrt{G}$.

When the two correlated modes are the two frequency sidebands at $\omega_0\pm\Omega$ modulated onto a single continuous-wave laser, Bob can measure the joint quadratures using a central LO at $\omega_0$ and a single BPD, followed by analog-to-digital conversion and digital I/Q demodulation \cite{su2023experimental,liao2025experimental}.
Let $\theta$ denote the relative phase between the signal and LO after proper phase locking.
The I/Q demodulation results of the photocurrent are proportional to $x_+\cos\theta+p_+\sin\theta$ and $-p_-\cos\theta+x_-\sin\theta$, respectively.
Bob switches the LO phase between $\pi/2$ and $0$ to select the amplified pair $(x_-, p_+)$ or the complementary pair $(x_+, p_-)$, and distills keys from the former.

For zero correlation, $c=0$, both quadrature pairs can be used to generate keys, giving the sifting probability $p_{\rm key}=1$, and the coherent-state protocol becomes equivalent to two parallel GG02 links.
In this case, Bob switches the PSA pump phase along with the LO phase so that each measured quadrature pair is amplified with the same gain and noise figure.
For nonzero correlations, Bob fixes the PSA pump phase to amplify the low-variance quadrature pair only as described above.

Figure~\ref{fig:protocol}(a) shows the complete receiver consisting of one two-mode PSA that amplifies the chosen key quadrature pair, and one BPD for joint quadratures readout.
One may argue that the correlated source encoding is not truly necessary since we can place the mode mixer $B$ at Bob before the joint receiver, as in Fig.~\ref{fig:protocol}(b).
This holds for identical independent thermal-loss channels and ideal mode mixing, but it need not hold for general multimode channels, which require dedicated effort for the security proof and protocol design.
But even for iid channels, a practical distinction remains.
Direct phase-conjugate encoding occurs at the transmitter, avoiding additional receiver-side mode conversion and its potential insertion loss and calibration requirements.

We characterize a practical PSA by the same net power gain $G$ and coherent-input linear noise figure $F$ for both key quadratures $q\in\{x_-,p_+\}$.
These parameters include the PSA module's insertion loss but exclude
the BPD noise.
Assuming linear amplification, the effective quadrature model for both preparation schemes can be written as
\begin{equation}
q_{\rm out}=\sqrt G\,q_{\rm in}+z,\qquad
\operatorname{Var}(z)=G(F-1)
\ ,
\label{eq:module}
\end{equation}
where $F-1$ is the PSA added-noise variance referred to its input.
Using a trusted receiver model and denoting the BPD efficiency by $\eta$ and electrical noise variance $v_{\rm el}$ \cite{usenko2016trusted}, the total receiver noise referred to the PSA input is given by
\begin{equation}
\chi_{\rm rec}
= F-1
+\frac{1-\eta+v_{\rm el}}{\eta G}
.
\label{eq:receiver_noise}
\end{equation}
An ideal PSA has $F=1$ and suppresses the detector contribution by $1/G$ \cite{fossier2009improvement,liao2025experimental}.
The complementary quadrature is assumed to be independently
calibrated for parameter estimation, and the required channel
covariances are taken as known in the asymptotic analysis below.

Each physical mode passes through an identical independent thermal-loss channel,
\begin{equation}
\begin{gathered}
\Gamma_B=T
(\Gamma_{S_1,S_2}+\chi_{\rm line}I_4)
\ ,
\end{gathered}
\label{eq:channel}
\end{equation}
where $T$ is the transmittance, $\chi_{\rm line}=(1-T)/T+\xi$ with excess noise $\xi$, and $\Gamma_B$ is the signal CM at the receiver input, before the PSA.

The asymptotic key rate per mode is \cite{devetak2005distillation,navascues2006optimality,garciapatron2006unconditional}
\begin{equation}
K = p_{\rm key}
\frac12
\max\{\beta (I_{AB}^{x_-} + I_{AB}^{p_+})-\chi_{BE}^{x_-,p_+},0\}
,
\label{eq:rate}
\end{equation}
where $p_{\rm key}$ is the probability of selecting the quadratures $\{x_-, p_+\}$, and $\beta$ is the reconciliation efficiency.
For both preparations, the mutual information satisfies $I_{AB}^{x_-} = I_{AB}^{p_+}$, and is found to be
\begin{equation}
\begin{gathered}
I_{AB}
=\frac12
\log_2
\left[ 1 +
\frac{\varepsilon}
{s+\chi_{\rm line}+\chi_{\rm rec}/T}
\right]
,
\end{gathered}
\label{eq:mutual}
\end{equation}
where $\varepsilon = V-c-s$ with the squeezed quadrature variance $s$, and $s=1$ for coherent preparation.
Appendix~\ref{app:rate} details the calculation of Holevo information.

\begin{figure*}[!t]
\centering
\includegraphics[width=\textwidth]{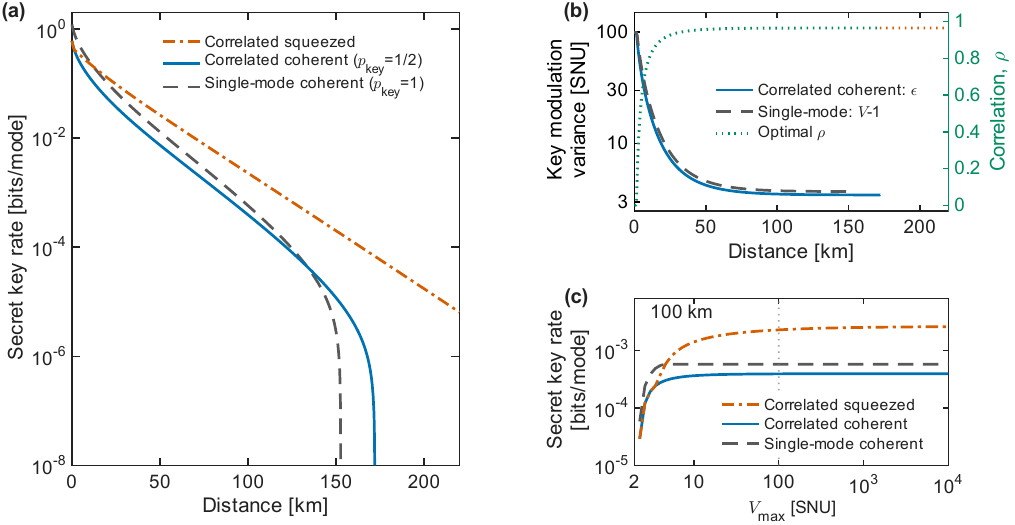}
\caption{Optimized secret key rate based on phase-conjugate encoding and joint reception.
(a) Secret key rates for correlated squeezed-state and coherent-state preparations and the single-mode coherent-state benchmark under $V\leq100$.
The correlated coherent encoding jointly optimizes $(V,c)$, while the single-mode benchmark optimizes $V$ with symmetric modulation.
The squeezed-state preparation uses the same optimized $(V,c)$ as the correlated coherent encoding.
Both correlated preparations use only the low-variance quadratures for key distribution ($p_{\rm key}=1/2$), whereas the single-mode benchmark uses both ($p_{\rm key}=1$).
(b) Left axis: key modulation variances $\varepsilon=V-c-1$ and $V-1$ for the correlated and single-mode coherent encodings, respectively.
Right axis: the optimal correlation $\rho$ (green dotted).
The orange dotted continuation denotes the squeezed-state preparation using the same optimized parameters as the coherent preparation.
(c) Secret key rates at 100 km versus the maximal $V$ used for optimization, denoted by $V_{\max}$.
The vertical dotted line marks $V_{\max}=100$, used in (a) and (b).
Other parameters are given in Table~\ref{tab:numerical_parameters}.}
\label{fig:optimized}
\end{figure*}

To allow correlations between the physical channels, we assume symmetric channels with common transmittance $T$ and excess noise $\xi$, and use a two-mode entangling cloner with input state characterized by \cite{ottaviani2017gaussian}
\begin{equation}
\Gamma_{E_1E_2}=
\begin{pmatrix}
\omega I_2&C_E\\
C_E&\omega I_2
\end{pmatrix},
\qquad
C_E=\operatorname{diag}(g_1,g_2)
\ ,
\label{eq:correlated_ancilla}
\end{equation}
where $\omega=1+T\xi/(1-T)$ is the ancillary mode variance.
The allowed correlation is determined by
\begin{equation}
\begin{gathered}
\omega\pm g_1>0,\qquad \omega\pm g_2>0,\\
(\omega+g_1)(\omega+g_2)\geq1,\\
(\omega-g_1)(\omega-g_2)\geq1.
\end{gathered}
\label{eq:attack_physicality}
\end{equation}
The channel output is now
\begin{equation}
\Gamma_B=T\Gamma_{S_1,S_2}+(1-T)\Gamma_{E_1E_2}
\ ,
\label{eq:correlated_channel}
\end{equation}
and the total added noise at the PSA input for quadrature $\{x_-, p_+\}$ is
\begin{equation}
\begin{aligned}
N_1&=(1-T)(\omega-g_1)+\chi_{\rm rec},\\
N_2&=(1-T)(\omega+g_2)+\chi_{\rm rec}.
\end{aligned}
\label{eq:correlated_key_noise}
\end{equation}

\section{Numerical Results and Discussion}
\label{sec:results}

\begin{table}[!t]
\caption{Default parameters used in the numerical studies unless stated otherwise.}
\label{tab:numerical_parameters}
\centering
\small
\setlength{\tabcolsep}{3pt}
\renewcommand{\arraystretch}{1.15}
\begin{tabular*}{\columnwidth}{@{\extracolsep{\fill}}ll@{}}
\toprule
Parameter & Default setting \\
\midrule
Fiber loss coefficient & $0.2$ dB/km \\
Input-referred excess noise & $\xi=0.05$ SNU \\
Reconciliation efficiency & $\beta=0.95$ \\
BPD efficiency & $\eta=0.72$ \\
Electronic-noise variance & $v_{\rm el}=1.19$ SNU \\
PSA net power gain & $G=10$ dB \\
PSA noise figure & NF $=1.3$ dB \\
Key-setting probability & $p_{\rm key}=1/2$ \\
\bottomrule
\end{tabular*}
\end{table}

\begin{figure*}[!t]
\begin{minipage}[t]{0.48\textwidth}
\vspace{0pt}
\centering
\includegraphics[width=\linewidth]{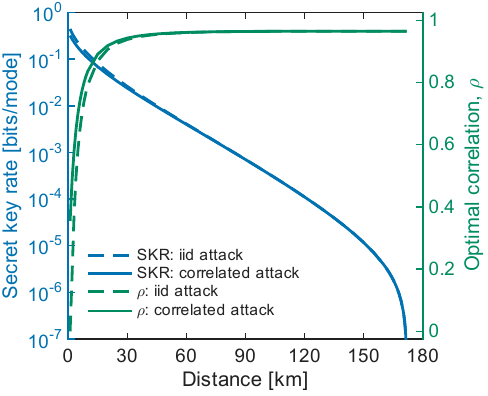}
\caption{Correlated coherent-state encoding under iid and correlated two-mode Gaussian attacks.
Blue curves give the secret key rate (left axis), and green curves give the optimal source correlation $\rho$ (right axis).
Dashed curves use $g_1=g_2=0$; solid curves minimize the rate over the attack domain in Eq.~\eqref{eq:attack_physicality}.
The source parameters $(V,c)$ are optimized separately for each attack model under $V\leq100$.
Both cases use $p_{\rm key}=1/2$ and the parameters in Table~\ref{tab:numerical_parameters}.}
\label{fig:correlated_performance}
\end{minipage}\hfill
\begin{minipage}[t]{0.48\textwidth}
\vspace{0pt}
\centering
\includegraphics[width=\linewidth]{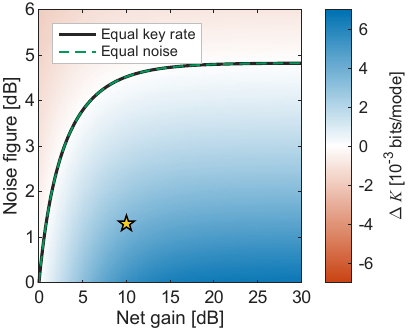}
\caption{Key-rate change relative to the case without using a PSA for correlated coherent-state encoding at 50 km, with $V=54$, $c=49.5$, $\eta=0.72$, and $v_{\rm el}=1.19$ SNU.
The black solid curve marks $\Delta K=0$, and the green dashed curve gives equality in Eq.~\eqref{eq:noise_threshold}; the two boundaries coincide within the scan resolution.
The star marks a net gain of 10 dB and NF $=1.3$ dB.
The source remains fixed, with $\xi=0.05$, $\beta=0.95$, and $p_{\rm key}=1/2$; all receiver noise is trusted.}
\label{fig:psa_gain_noise}
\end{minipage}
\end{figure*}

\subsection{Optimized phase-conjugate encoding}
\label{sec:optimized_encoding}

Using the normalized correlation defined by \eqref{eq:rho} and the parameters outlined in Table~\ref{tab:numerical_parameters}, Fig.~\ref{fig:optimized}(a) compares the phase-conjugate encoded coherent-state and squeezed-state protocols with a standard single-mode coherent-state benchmark (GG02).
For coherent encoding, the ensemble variance $V \le 100$ and correlation $\rho$ are jointly optimized at each distance, whereas the single-mode benchmark uses symmetric modulation and optimizes only $V$.
The probability of selecting the joint quadratures $\{x_-, p_+\}$ is $p_{\rm key} = 1/2$ in both correlated protocols, whereas the single-mode case has $p_{\rm key} = 1$.
The squeezed-state encoding uses the same optimized parameters $(V,c)$ as the coherent encoding and improves performance because of a higher $I_{AB}$.
At 100 km it requires 8.17 dB of intrinsic squeezing and gives $2.26\times10^{-3}$ bits/mode.
Correlated coherent encoding extends the maximum distance of the single-mode benchmark from 152.6 to 171.9 km with the optimal correlation changing from $\rho=0$ at zero distance to $\rho\simeq0.965$ at the maximum distance, as shown in Fig.~\ref{fig:optimized}(b).
Although the single-mode protocol gives higher rates at short and intermediate distances, the correlated coherent encoding overtakes it at approximately 133.3 km.
As discussed in Sec.~\ref{sec:encoding_boundaries}, the correlated coherent preparation reduces to GG02 in the boundary case with $\rho=0$ and $p_{\rm key}=1$.
This suggests that the best practice with the proposed two-mode coherent-state CVQKD is using zero correlations and $p_{\rm key}=1$ for distances below the performance crossing point and using strong correlations and $p_{\rm key}=1/2$ for distances beyond.

The optimized $(V,\rho)$ also determines the key modulation variance $\varepsilon = V-c-1$ of the correlated coherent encoding, which is compared with the single-mode modulation variance $V_A = V-1$ in Fig.~\ref{fig:optimized}(b).
The optimal modulation variances approach finite values in both cases and strong correlation is needed for the correlated protocols.
Figure~\ref{fig:optimized}(c) further confirms the performance of the correlated coherent protocol is insensitive to the maximal $V$ beyond 100 used above, checked at 100 km.
The squeezed-state preparation benefits slightly from the larger $V$, at the cost of a larger intrinsic squeezing.

Equation~\eqref{eq:coherent_key_records} shows that $X$ and $P$ cancel in the low-variance records and appear only in the high-variance quadratures.
For the independent channels in \eqref{eq:channel}, varying $X$ and $P$ therefore does not change the mutual information $I_{AB}$ in the quadratures $(x_-,p_+)$, as in \eqref{eq:mutual}, but provides randomization of the complementary quadratures $(x_+,p_-)$.
This is equivalent to applying an additional random-displacement channel to Eve's conditional states without changing Bob's key statistics.
The data-processing inequality then implies that $\chi_{BE}$ cannot increase under this operation \cite{lindblad1975completely}.
Numerical checks at fixed modulation level $\varepsilon$ confirm that $\chi_{BE}$ decreases and approaches saturation as the variances of $X$ and $P$ increase.
A closely related effect was identified for trusted preparation noise in squeezed-state CVQKD, where noise in the anti-squeezing quadrature can reduce Eve's Holevo information without changing $I_{AB}$ \cite{oruganti2025noisy}.
In the decoupled mode basis, our additional private modulation plays a similar role in the high-variance quadrature of each mode.

\subsection{Performance under correlated attacks}
\label{sec:correlated_results}

We next compare the correlated coherent-state protocol under iid and correlated two-mode Gaussian attacks, using the model in \eqref{eq:correlated_ancilla}.
At each distance, Alice optimizes $(V,c)$ separately for the two attack models under $V\leq100$, as in Fig.~\ref{fig:optimized}.
The iid reference fixes $g_1=g_2=0$.
For correlated attacks, Eve minimizes the key rate over $(g_1,g_2)$ within the physical domain in \eqref{eq:attack_physicality}.

Figure~\ref{fig:correlated_performance} shows that allowing correlations in Eve's ancillary modes lowers the optimized key rate.
Compared with the iid result, the rate decreases by 17.9\% at 10 km, 3.32\% at 50 km, and 0.387\% at 100 km.
The results also show that the preference for stronger source correlations at longer distances persists under correlated attack.
Numerical scans at various distances, with the corresponding optimized source parameters held fixed, support that the optimal attack satisfies $g_1=-g_2>0$ (see Fig.~\ref{fig:attack_landscape} in Appendix~\ref{app:controls}).
In the decoupled mode basis, these correlations represent phase-sensitive environmental noise, with opposite orientations in the two decoupled modes.

\subsection{PSA compensation of detector noise}

According to \eqref{eq:receiver_noise}, a practical PSA suppresses the detector noise contribution by $1/G$ but also adds noise.
Let the detector noise be $\chi_0=(1-\eta+v_{\rm el})/\eta$; then the total receiver noise is $\chi_{\rm rec}=F-1+\chi_0/G$.
The PSA reduces the total receiver noise when
\begin{equation}
F<1+\chi_0(1-G^{-1}).
\label{eq:noise_threshold}
\end{equation}
For large enough gain, the total receiver noise consists of mainly the PSA added noise since $\chi_{\rm rec} \to F-1$ and the PSA benefits only when its NF satisfies $F<1+\chi_0$.
Lower NF is therefore important once the detector contribution has been sufficiently suppressed.

Figure~\ref{fig:psa_gain_noise} examines the combined effect of gain and NF for the coherent-state preparation with the detector considered here, with $\eta=0.72$ and $v_{\rm el}=1.19$ SNU.
The color scale shows the rate change $\Delta K=K(G,F)-K(1,1)$ relative to the no-PSA case, keeping the source and channel unchanged.
For $V=54$ and $c=49.5$ at 50 km, the region with $\Delta K>0$ agrees with the noise-reduction condition in \eqref{eq:noise_threshold}.
The 10 dB, 1.3 dB-NF module increases the key rate from $3.81\times10^{-3}$ to $7.26\times10^{-3}$ bits/mode.

To isolate the PSA effect on the detector loss and noise, we can examine the NF requirement in \eqref{eq:noise_threshold} by setting $v_{\rm el}=0$ or $\eta=1$, respectively.
For loss alone at $\eta=0.72$, a net gain of 10 dB requires NF below 1.303 dB to satisfy this condition.
The 1.3 dB-NF considered here therefore gives almost no net improvement in terms of compensating the detector loss.
For the noisy detector with $\eta=1$, the NF threshold is 3.162 dB and for both $\eta=0.72$ and $v_{\rm el}=1.19$ SNU, the NF threshold increases to 4.529 dB at the same 10 dB gain.

\begin{figure}[!t]
\centering
\includegraphics[width=\columnwidth]{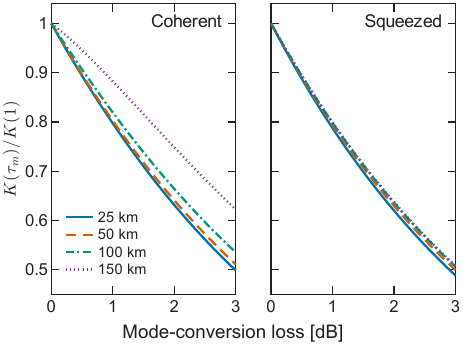}
\caption{Effect of receiver-side mode-conversion loss on coherent-state (left) and squeezed-state (right) preparations of the same average state.
Rates are normalized to their zero-loss values.
The source is optimized for coherent preparation without the converter under $V\leq100$, using the same iid calculation as Fig.~\ref{fig:optimized}, and then held fixed for both preparations.
The PSA has a net gain of 10 dB and NF $=1.3$ dB, with $\eta=0.72$, $v_{\rm el}=1.19$ SNU.}
\label{fig:conversion}
\end{figure}

\subsection{Receiver-side mode conversion}

Direct generation of the correlated modes avoids the receiver-side mode converter in Fig.~\ref{fig:protocol}(b).
To see the impact of a lossy converter, we model it as ideal mixing followed by equal trusted losses of transmittance $\tau_m$ in the two modes.
The combined converter and PSA then have $G'=\tau_mG$ and $F'={F}/{\tau_m}$, with total receiver noise referred to the converter input
\begin{equation}
\chi'_{\rm rec}
=\frac{1-\tau_m+\chi_{\rm rec}}{\tau_m}.
\label{eq:converter_noise}
\end{equation}

Figure~\ref{fig:conversion} shows the resulting key-rate reduction for both coherent and squeezed preparations.
At each distance, the coherent source is optimized under the same condition as in Fig.~\ref{fig:optimized}, and its parameters are then fixed while the conversion loss is varied.
An additional loss of 1 dB reduces the coherent-state rate by 19.7\% at 50 km and 18.1\% at 100 km, with similar reductions for squeezed-state preparation.

Note that the source--receiver equivalence used for this conversion comparison is restricted to iid channels; unequal or generally correlated multimode channels need not commute with the mode mixer.

\section{Conclusion}

We have developed a tunable phase-conjugate Gaussian encoding and security framework for two-mode CVQKD with a single PSA and joint detection using one BPD.
We present both coherent-state and squeezed-state protocols based on the same two-mode source with tunable intermode correlations.
The encoding and detection scheme separates joint quadratures used for key generation from complementary quadratures carrying private random modulation, which can reduce Eve's information without changing the Alice--Bob mutual information under the channel models considered here.
The correlated coherent-state encoding gives higher rates at long distances compared to the single-mode coherent-state benchmark and extends the maximal distance under the same receiver-noise conditions.
The preferred intermode correlation increases with transmission distance, and the distance advantage persists when the encoding is reoptimized against the symmetric correlated two-mode Gaussian attacks.
The practical benefit of correlated encoding is direct compatibility with the shared PSA and joint measurement.
The coherent preparation uses classically precoded Gaussian displacements of coherent sidebands, while the squeezed preparation applies correlated displacements to a directly generated two-mode squeezed vacuum.
Neither implementation requires an additional optical mode converter, preserving a rate advantage in a nonideal implementation with finite mode conversion loss.

\appendix

\section{The EB source and generalized heterodyne measurements}
\label{app:decompositions}

The EB source in Fig.~\ref{fig:common_eb}(a) starts from two independent pure EPR pairs.
Each pair has CM
\begin{equation}
\Gamma_{\rm EPR}(u)=
\begin{pmatrix}
uI_2&kZ\\
kZ&uI_2
\end{pmatrix},
\qquad k=\sqrt{u^2-1}.
\label{eq:epr_source_encoding}
\end{equation}
After the opposite squeezing and mode mixing defined in Sec.~\ref{sec:source}, the four-mode CM in the ordering $(A_1,A_2,S_1,S_2)$ can be written as
\begin{equation}
\begin{gathered}
\Gamma_{\rm EB}=
\begin{pmatrix}
uI_2&0&C_0/\sqrt2&-C_0/\sqrt2\\
0&uI_2&\widetilde C_0/\sqrt2&\widetilde C_0/\sqrt2\\
C_0^T/\sqrt2&\widetilde C_0^T/\sqrt2&VI_2&cZ\\
-C_0^T/\sqrt2&\widetilde C_0^T/\sqrt2&cZ&VI_2
\end{pmatrix}, \\
C_0=kZS_r
=k\operatorname{diag}(\sqrt{s},-1/\sqrt{s}), \\
\widetilde C_0=kZS_{-r}
=k\operatorname{diag}(1/\sqrt{s},-\sqrt{s}),
\end{gathered}
\label{eq:full_four_mode_source}
\end{equation}
which is a purification of the average signal state $\Gamma_{S_1,S_2}$.

A generalized heterodyne measurement can be implemented by mixing the input mode with vacuum on a beam splitter of transmittance $\tau_A$, homodyning orthogonal output quadratures, and rescaling the outcomes to obtain unit gain for the input quadratures.
The equivalent measurement noise CM is
\begin{equation}
\begin{gathered}
M=\operatorname{diag}\left(
m,m^{-1}
\right), \qquad
m = \frac{1-\tau_A}{\tau_A}
\end{gathered}
\label{eq:encoding_generalized_heterodyne}
\end{equation}

Applying this measurement to $A_1$ in Eq.~\eqref{eq:full_four_mode_source}, the conditional CM of $b_1$ before mode mixing is \cite{genoni2016conditional}
\begin{equation}
\begin{aligned}
\gamma_{b_1|a_1}
&=\gamma_{b_1}-C_0^T(uI_2+M)^{-1}C_0\\
&=\operatorname{diag}(us,u/s)-C_0^T(uI_2+M)^{-1}C_0\\
&=\operatorname{diag}(v_m,v_m^{-1}),
\end{aligned}
\end{equation}
where
\begin{equation}
\gamma_{b_1}=S_r(uI_2)S_r^T = \operatorname{diag}(us,u/s)
\end{equation}
describes the $b_1$ mode before Alice's measurement, and
\begin{equation}
v_m=s\frac{um+1}{u+m}.
\end{equation}
Requiring $v_m=1$ gives
\begin{equation}
m=\frac{u-s}{us-1}=\frac{u-s}{V-c-1},
\end{equation}
which agrees with Eq.~\eqref{eq:coherent_seed}.
We have used $us=V-c$ and $u/s=V+c$ for $m$.
These relations follow from the source parameters defined in Sec.~\ref{sec:source},
\begin{equation}
u=\sqrt{(V-c)(V+c)} \ ,
\quad
s=e^{-2r}=\sqrt{\frac{V-c}{V+c}}
\ .
\end{equation}
The measurement on $A_2$ likewise gives $\gamma_{b_2|a_2}=I_2$.
Thus the conditional state is a product of coherent states, before and after the passive mixing $B$.

\section{Holevo information with trusted receiver noise}
\label{app:rate}

Model Eve's attack by a two-mode entangling cloner that injects modes $e=(E_1,E_2)$ with CM $\Gamma_e=\Gamma_{E_1E_2}$ in Eq.~\eqref{eq:correlated_ancilla}, and correlations satisfying Eq.~\eqref{eq:attack_physicality}.
Eve couples each injected mode to the corresponding signal mode through a beam splitter of transmittance $T$ and keeps both reflected outputs and the modes purifying $\Gamma_e$.
These modes together form her system $E$.

Let $A=(A_1,A_2)$ denote Alice's modes and $B=(B_1,B_2)$ Bob's at channel output but before the receiver.
From the EB source in Eq.~\eqref{eq:full_four_mode_source}, their CM is
\begin{equation}
\begin{gathered}
\Gamma_{AB}=
\begin{pmatrix}
\Gamma_A&C_{AB}\\
C_{AB}^T&\Gamma_B
\end{pmatrix},\qquad \Gamma_A=uI_4,\\
\Gamma_B=T\Gamma_S+(1-T)\Gamma_e,\qquad
C_{AB}=\sqrt T\,C_{AS},
\end{gathered}
\label{eq:app_ab}
\end{equation}
where $\Gamma_S=\Gamma_{S_1,S_2}$ and $C_{AS}$ is the cross-covariance part in Eq.~\eqref{eq:full_four_mode_source}.
Since Eve purifies $AB$, her unconditional entropy is $S(E)=S(AB)$.

For the channel model considered here, Bob's transformation $B^T$ would decouple the two correlated modes and, with proper local phase rotations and reordering, reduce $\Gamma_{AB}$ to two independent Gaussian blocks:
\begin{equation}
\begin{gathered}
\gamma_j=
\begin{pmatrix}
uI_2&\sqrt T\,C_0\\
\sqrt T\,C_0^T&\gamma_{B,j}
\end{pmatrix},\qquad j=1,2,\\
\gamma_{B,j}=T\operatorname{diag}(V-c,V+c)+(1-T)D_j,
\end{gathered}
\label{eq:app_channel_blocks}
\end{equation}
where $C_0$ is defined in Eq.~\eqref{eq:full_four_mode_source} and
\begin{equation}
\begin{gathered}
D_1=\operatorname{diag}(\omega-g_1,\omega-g_2),\\
D_2=\operatorname{diag}(\omega+g_2,\omega+g_1).
\end{gathered}
\end{equation}

Given the total receiver noise $\chi_{\rm rec}$ in Eq.~\eqref{eq:receiver_noise} and $d=(1+\chi_{\rm rec})^{-1}$, the CM used for conditioning is \cite{yamano2023general}
\begin{equation}
\begin{gathered}
\mathcal M_j=\mathcal R_d(\gamma_j\oplus I_2)\mathcal R_d^T,\\
\mathcal R_d=I_2\oplus
\begin{pmatrix}
\sqrt{1-d}\,I_2&-\sqrt d\,I_2\\
\sqrt d\,I_2&\sqrt{1-d}\,I_2
\end{pmatrix}.
\end{gathered}
\label{eq:app_receiver_cm}
\end{equation}
The mode ordering in $\mathcal M_j$ is $(A_j,F_j,Y_j)$, where $F_j$ is the unused trusted receiver mode and $Y_j$ is measured.
Writing
\begin{equation}
\mathcal M_j=
\begin{pmatrix}
\gamma_{A_jF_j}&K_j\\
K_j^T&\gamma_{Y_j}
\end{pmatrix},
\end{equation}
the homodyne conditional CM is \cite{weedbrook2012gaussian}
\begin{equation}
\widetilde\gamma_j
=\gamma_{A_jF_j}-K_j(\Pi\gamma_{Y_j}\Pi)^{+}K_j^T,
\label{eq:app_conditional_trusted}
\end{equation}
where $\Pi=\operatorname{diag}(1,0)$ and $+$ denotes the Moore--Penrose pseudoinverse.

Denote the two symplectic eigenvalues of $\gamma_j$ and $\widetilde\gamma_j$ as $\nu_{j,\pm}$ and $\widetilde\nu_{j,\pm}$.
The two-mode Holevo information is
\begin{equation}
    \begin{gathered}
\chi_{BE}^{x_-,p_+}
=\sum_{j=1}^2\sum_{\sigma=\pm}
\left[h(\nu_{j,\sigma})-h(\widetilde\nu_{j,\sigma})\right] \ ,\\
h(\nu)=\frac{\nu+1}{2}\log_2\frac{\nu+1}{2}
-\frac{\nu-1}{2}\log_2\frac{\nu-1}{2}
\ .
\end{gathered}
\label{eq:app_joint_holevo}
\end{equation}
The mode number normalization and the factor $p_{\rm key}$ are applied separately in Eq.~\eqref{eq:rate}.

\section{Correlated-attack optimization}
\label{app:controls}

\begin{figure}[!t]
\centering
\includegraphics[width=\columnwidth]{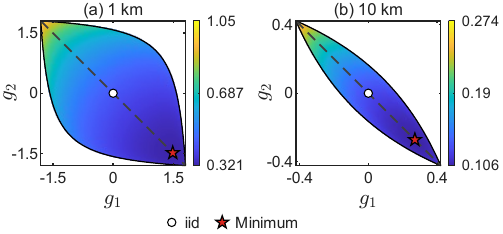}
\caption{Correlated attack scans at fixed optimized coherent-state sources.
Panels (a) and (b) show the key rates at 1 and 10 km, with $(V,c)=(100,34.98)$ and $(100,82.82)$, respectively.
Solid black curves mark the physical boundaries, dashed lines indicate $g_2=-g_1$, open circles mark the iid attack, and stars mark the refined minima.
All other parameters are those of Fig.~\ref{fig:correlated_performance}.}
\label{fig:attack_landscape}
\end{figure}

Figure~\ref{fig:attack_landscape} examines the correlated attack plane over the physically allowed region of $(g_1,g_2)$ at 1 and 10 km, fixing the corresponding source parameters from Fig.~\ref{fig:correlated_performance} at $(V,c)=(100,34.98)$ and $(100,82.82)$.
The color scale gives the rate $K_{\rm raw}=p_{\rm key}(\beta I_{AB}-\chi_{BE})$, without reoptimizing the source at each attack point.
The refined minima lie on $g_1=-g_2$, with $g_1\simeq1.48416$ and $0.271269$ at the two distances, respectively.

\bibliography{reference}
\end{document}